\documentclass[apj]{emulateapj}
\newcommand{\simgt}{\lower.5ex\hbox{$\; \buildrel > \over \sim \;$}}
\newcommand{\simlt}{\lower.5ex\hbox{$\; \buildrel < \over \sim \;$}}

\newcommand{\baredth}{\;\overline{\raise1.0pt\hbox{$'$}\hskip-6pt \partial}\;}
\newcommand{\edth}{\;\raise1.0pt\hbox{$'$}\hskip-6pt\partial\;}

\slugcomment{Received 2005 June; accepted 2005 July}

\shorttitle{Multiply Imaged Galaxies Behind SDSS~J1004+4112}

\shortauthors{SHARON ET Al.}

\begin{document}

\title{Discovery of  multiply imaged galaxies behind the cluster
and lensed quasar SDSS~J1004+4112
\footnotemark[1]}\footnotetext[1]{Based on 
observations made with the NASA/ESA {\it Hubble Space Telescope},
   which is operated by AURA, Inc., under NASA contract NAS5-26555.}

\author{Keren Sharon\altaffilmark{2},
Eran O. Ofek\altaffilmark{2}, 
Graham P. Smith\altaffilmark{3},
Tom J. Broadhurst\altaffilmark{2}, 
Dan Maoz\altaffilmark{2},
Christopher S. Kochanek\altaffilmark{4}, 
Masamune Oguri\altaffilmark{5,6}, 
Yasushi Suto\altaffilmark{5},
Naohisa Inada\altaffilmark{7}, 
Emilio E. Falco\altaffilmark{8}
}

\altaffiltext{2}{School of Physics and Astronomy, Tel Aviv University, Tel-Aviv 69978, Israel}
\altaffiltext{3}{Department of Astronomy, California Institute of Technology, Mail Code 105--24, Pasadena, CA 91125}
\altaffiltext{4}{Department of Astronomy, Ohio State University, Columbus, OH 43210}
\altaffiltext{5}{Department of Physics, School of Science, University of Tokyo, Tokyo 113-0033, Japan}
\altaffiltext{6}{Department of Astrophysical Science, Princeton University, Princeton, NJ 08544, USA}
\altaffiltext{7}{Institute of Astronomy, Faculty of Science, University of Tokyo, 2-21-1 Osawa, Mitaka, Tokyo 181-0015, Japan.}
\altaffiltext{8}{Harvard-Smithsonian Center for Astrophysics, 0 Garden St., Cambridge, MA 02138}

\begin{abstract}

We have identified three multiply imaged galaxies in {\it Hubble Space Telescope}
images of the redshift $z=0.68$ cluster responsible for the large-separation 
quadruply lensed quasar, SDSS~J1004+4112. Spectroscopic redshifts have been secured 
for two of these systems using the Keck~I 10m telescope. The most distant lensed galaxy, at
$z=3.332$, forms at least four
 images, and an Einstein ring encompassing $3.1$ times more area than the 
 Einstein ring of the lensed QSO images at $z=1.74$, due to the greater source
distance. For a second multiply imaged galaxy, we identify  Ly$\alpha$ emission at a redshift
of $z=2.74$.  The cluster mass profile can be constrained
from near the center of the brightest cluster galaxy, where we observe both a radial
arc and the fifth image of the lensed quasar, to the Einstein radius
of the highest redshift galaxy, $\sim 110$~kpc. Our preliminary
modeling indicates that the mass approximates
an elliptical body, with an average projected logarithmic  gradient of
$\simeq -0.5$.
The system is potentially useful for a direct measurement of world models in
a previously untested redshift range. 
\end{abstract}                   


\section{Introduction}\label{section1}

 The recent discovery of a quadruply lensed quasar, SDSS J1004+4112,
with unusually large image separations (Inada et al. 2003, Oguri et al. 2004),
has generated much interest.  The quasar, at $z=1.74$, is lensed by a galaxy
cluster at $z=0.68$ into four bright images on an 
Einstein ring of approximately $15''$ in diameter. A 
faint fifth image is seen projected through
the inner isophotes of the central brightest cluster galaxy (BCG; Inada et al. 2005). 
The QSO is known to vary in brightness. Assuming a 
nominal value for the Hubble parameter 
$H_0$, it will be possible, for the first time, to model a cluster
potential using complementary information on the value
of the potential itself (from the measured time delays), 
rather than only constraints
on the first derivative of the potential (i.e., multiple-image positions)
and the second derivative of the potential (i.e., weak lensing).
Finally, the partial Einstein ring, seen to pass through the
lensed quasar images, is the most highly magnified quasar host galaxy
known, permitting a unique probe of quasar hosts at $z=1.7$ (Kochanek
et al. 2005, in prep.). 

 Strong lensing is now being used to constrain in detail the inner
mass profile of galaxy clusters and for estimating the amount of 
sub-structure (e.g., Gavazzi et al. 2004, Kneib et al. 2003, Smith et
al. 2005, Sand et al. 2004, Broadhurst et al. 2005a,b). A 
prediction of N-body simulations using the standard $\Lambda$-CDM
cosmology (e.g., Navarro, Frenk, \& White 1997, NFW; Moore et al. 1999) 
is that the logarithmic gradient of the density profile of the cluster
should be shallower than that of an isothermal profile.
In the case of A1689, where 30 background galaxies are found to be
lensed into over 100 images (Broadhurst et al. 2005a), the profile 
does continuously flatten towards the center like an NFW profile, but
with a surprisingly high concentration, $c_{vir}=14\pm1.5$ (Broadhurst et
al. 2005b), compared with the much more diffuse halos predicted for
massive clusters, $c_{vir}\sim4$ (e.g., Bullock et al.  2001). A1689 has the
largest known Einstein ring, $\theta_E=48''$ (for $z=3$) and thus
projection effects may be important (Oguri et al. 2005).
A survey of more typical clusters is needed before drawing any general
conclusions.

The Advanced Camera for Surveys (ACS) on the 
{\it Hubble Space Telescope} (HST) has an
unparalleled advantage for lensing work, providing depth, spatial
resolution and color information.  Well resolved
internal structure can be helpful in identifying counter-images
of multiply imaged galaxies, particularly since distant star-forming
galaxies often have complex and unique morphologies. A good lens model
is required in order to take proper advantage of the resolved internal
structure, as the parity and differential magnification of internal
galaxy features varies between images of the same source, leading to
confusion. 
Without guidance from a lens model, it
is hard to recognize counter-images because they often fall in
unexpected places due to deflections by substructure and cluster
galaxies.
 The angular position, $\vec{\theta}_i$, of an image is given by the
 lens equation $\vec{\theta}_i=\vec{\beta}_s +
 {(d_{ls}/{d_{s}})}\vec{\alpha}(\theta_i)$, where
 $\vec{\alpha}(\theta_i)$ is the deflection field of the lens, 
 $\vec{\beta}_s$ is the location of the source in the source plane
 and 
$d_{ls}/d_s$ is the ratio of angular diameter distances from the lens to the source 
and from the observer to the source, respectively.
Thus, predictions must be made for a wide range of background source distances unless the
source redshift is known.

It should be appreciated that all the images of background galaxies
lying within approximately twice the Einstein radii 
pertaining to their redshifts 
will have one or more lensed counter-images. Hence the usual identification
of, at most, a few sets of multiple images per cluster, even in
deep HST images, means the ``eyeball'' approach to finding
counter-images can be far from exhaustive. We have had more success by
using an iterative method, which we describe in \S2.

Here we report the identification of three new examples of
multiply imaged galaxies behind the cluster SDSS~J1004+4112, and the
measurement of spectroscopic redshifts for two of these. We use the
multiply imaged galaxies and the five images of the QSO for a
preliminary estimate of the
cluster mass distribution. In a future paper (Ofek et al. 2005, in
preparation), we will study more exhaustively the joint constraints that
can be imposed on both the mass distribution and on the background cosmology.
Throughout this paper we will assume cosmological parameters 
$H_0=70$ km s$^{-1}$ Mpc$^{-1}$, $\Omega_m=0.3$, and $\Omega_\Lambda=0.7$.

\section{Identification of multiple images} 
\label{multple}

We have examined the single-orbit 
ACS Wide Field Camera
images of SDSS~J1004+4112 taken
    in $I$(F814W) in 2004, April 28, 
and in $V$(F555W) in 2005, January 27, each 
with a total exposure time of 2025~s, 
 obtained as part of the CASTLES program
(GO-9744, Falco et al. 2001). Individual exposures in each band were processed
 and combined with the standard Space Telescope Science Institute
 pipeline. In addition to the final images in each band, we have
 produced a ``true-color'' image by assigning a blue color to the $V$ count
 rate, red to the $I$ count rate, and green to a
 linear interpolation of the $V$ and $I$ count-rates (see Fig. 1).
 Our goal is to identify multiple-image systems across as wide a range
of redshifts as possible.
In these data, we have identified three sets of
multiply imaged galaxies, in addition to the known images of the
lensed QSO.  

The identification of multiply imaged galaxies is achieved in an
iterative manner. We first construct a simplified mass model, as
described in \S4, initially based only on the images of the QSO. 
We de-lens the pixels belonging to any given faint object, using 
the lens model, onto a sequence of source planes of different distances,
re-lens these pixels back to the image plane, and compare the
model-predicted locations and morphology of the new counter-images with
the data. 
As new images are identified, they are incorporated into a refined lens model,
enhancing the prospects of finding further sets of multiply imaged galaxies.

The first set of lensed galaxy images to be identified in this
way are labeled Galaxy A (A1-A5 in Figs.~1 and 2). They are observed as three arc-like
images (A2, A3, A4) that lie well outside the Einstein ring traced by the
QSO images.
As seen in Fig.~1, the distance scale increases by a large
factor between the redshift of the quasar and of Galaxy A. 
 A counter-image of this galaxy is predicted by the
model and readily identified as A1, with an accurate reproduction of the
internal morphology. 
A radial feature observed close to
the central BCG (best seen in Fig.~2) could be a fifth counter-image of  
galaxy A. However, depending on the exact slope of the inner region 
of the mass profile, it could also be associated with other galaxies
in the field. 
Deeper multi-color observations are required to
confirm this possible association.
This radial arc together with the
fifth de-magnified image of the quasar will be helpful in
constraining the inner mass distribution.  

Two other sets of
multiply imaged galaxies, Galaxies B and C, 
 were identified using the best fit
 model based on galaxy A and the QSO.  Each is identified as a close
 pair of images (see Figs. 1 and 2).
Galaxy C has a different angular deflection
scale, and therefore must lie at a higher redshift than galaxy B
(see Table 1 for predictions). The model also predicts that galaxies B and
C should both have faint, de-magnified images that are buried in the
high surface brightness BCG and hence are undetectable in the
available data.

\begin{figure*}
\resizebox{170mm}{!}{\plotone{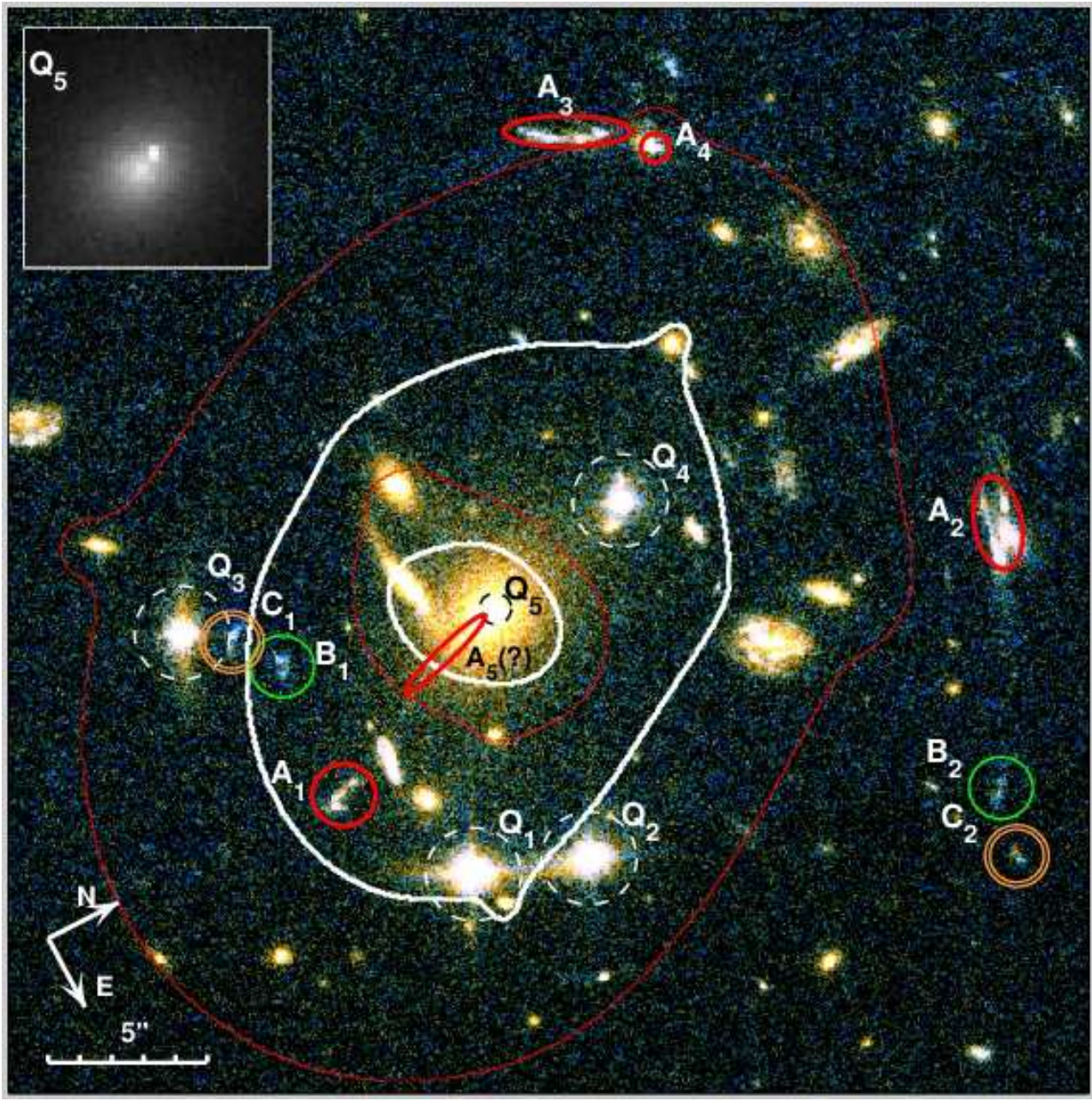}}
\caption{A color image of SDSS~J1004+4112
produced from the ACS $V$- and $I$-band observations,
with the images of the lensed galaxies and the QSO indicated. 
The inset shows the central region in $I$, where
a faint fifth image of
the lensed QSO is visible close to the center of the BCG. 
The fifth image is detected in both $V$ and $I$, 
and has a $V$-$I$ color consistent with those of the other quasar images.
Overlaid are the critical curves (radial and tangential),
based on our model for the mass distribution, for
the lensed QSO (white), and the corresponding critical curves for
Galaxy A (red). Notice that the area enclosed by  the Galaxy A 
tangential critical curve
(essentially, the Einstein ring) 
is much larger ($\times 3.1$) than that of the QSO,
due to the greater distance of Galaxy A ($z=3.332$ vs. $z=1.74$).
The Vega-based $I$ magnitudes of the galaxy images are: A1 23.4; A2 22.3; A3 22.4; A4 23.8;
A5 26.6; B1 24.7; B2 25.0; C1 24.1; C2 24.8.}
\end{figure*}

\begin{figure*}
\resizebox{170mm}{!}{\plotone{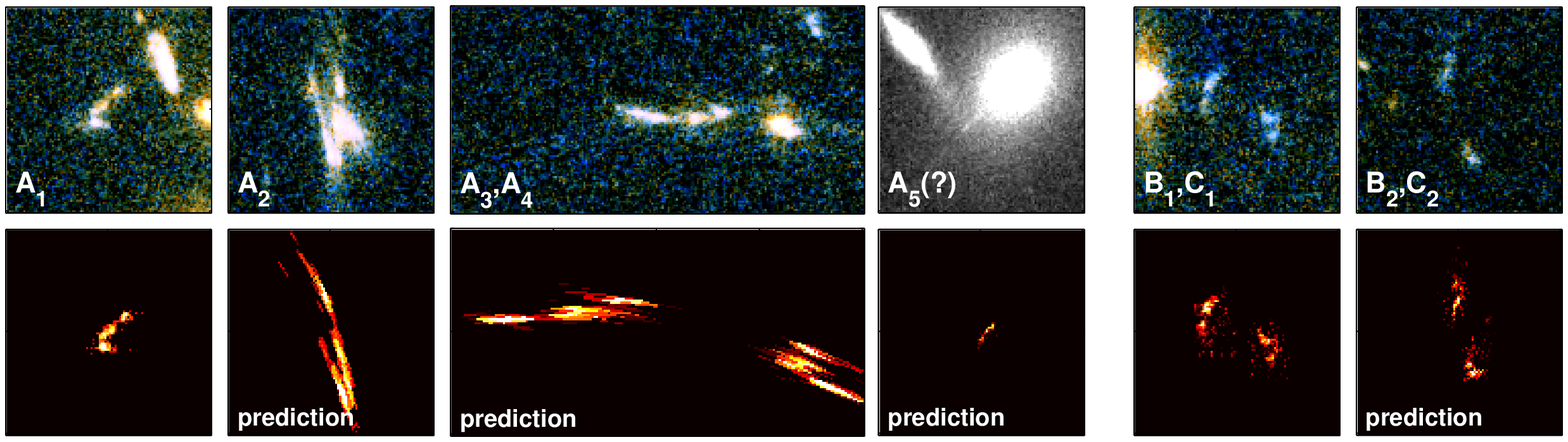}}
\caption{$V+I$-based color renditions of the observed lensed sources
  (upper row)
compared with the corresponding model-generated images in the lower
row. The left-most image of each source is the template used to predict the
structure of the other counter-images.}
\end{figure*}

\section{Spectroscopy}

 Multi-slit spectroscopy was carried out with LRIS on
 the Keck I 10~m telescope, on the nights of 2005, March 4 and 6, totaling 
 2.2~hrs of exposure time
 in dark conditions with $1''$ seeing. 
The D560 dichroic was used with the 400/3400 grism and 600/7500 grating
at 6850~\AA, thus providing continuous wavelength coverage of 
3500~\AA$\simlt\lambda_{obs}\simlt7900$~\AA.
The FWHM spectral resolution and pixel scales were 6.8~\AA, 4.5~\AA~ and
$0\farcs214$ pixel$^{-1}$, $0\farcs135$ pixel$^{-1}$ on the blue and red cameras, respectively.

Spectra were taken of the A2, A3 and A4 images of galaxy A, image B1 of
 galaxy B and image C1 of galaxy C.

\noindent{\bf Galaxy A} -- The one dimensional spectra extracted for A2, A3, and A4 
(Fig.~3) clearly show that these are images of the same galaxy, although with 
some contamination of foreground galaxy light to A2 and A4. 
Several interstellar absorption lines were identified,
from which we derive a redshift of $z=3.332$ for Galaxy A.

\noindent{\bf Galaxy B} -- In galaxy image B1 (Fig.~1) we identified a
single emission line at $\lambda_{\rm obs}{=}4547$~\AA, with an equivalent
width of $80$~\AA~ and intrinsic FWHM of $310~{\rm km\,s^{-1}}$.  We interpret
this line as Ly${\alpha}$ redshifted to $z{=}2.74$, excluding the lower
redshift alternative ([OII] at $z{=}0.22$, i.e., in front of the cluster)
on the basis of the presence of multiple-images of this galaxy and the
absence of H${\beta}$ and [OIII] in the red-arm spectrum.
Since the instrumental resolution and the intrinsic width of the line are
comparable, the Ly$\alpha$ profile appears relatively symmetric.

\noindent{\bf Galaxy C} -- We detected a weak continuum in C1, 
but no obvious spectral features.  The redshift of Galaxy C
therefore remains unconstrained spectroscopically.

In summary, we have measured the redshifts of two of the multiply imaged
galaxies behind this cluster, the spectroscopic redshifts comparing 
favorably with the redshifts inferred from a lens model based only on the image positions and 
the quasar redshift.

\begin{figure}[t]
\begin{center}
\resizebox{9cm}{!}{\plotone{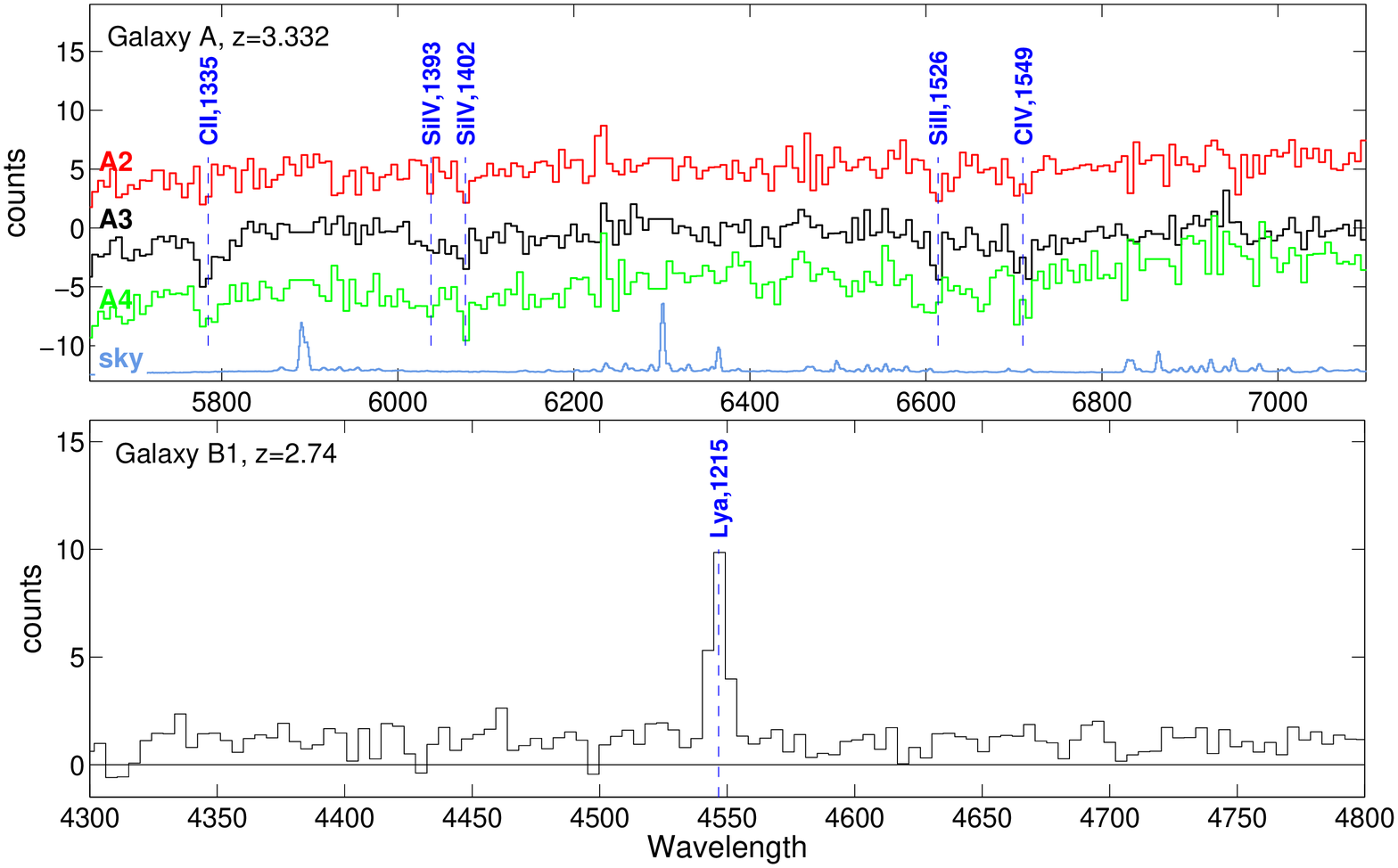}}
\end{center}
\caption{Top panel: 
Keck spectra (red arm) of images A2, A3 and A4 of Galaxy A, with
identified spectral features for the 
derived redshift  of $z=3.332$. The A3 and A4 spectra have been
vertically shifted for clarity. Sky subtraction residuals at 
5892~\AA, 6300~\AA, and 6864~\AA~ have been excised from the spectra --
a sky spectrum, showing the main atmospheric
emission lines, is shown for comparison. 
Slight differences in the spectra of the three galaxies are due to
contamination of the spectra of A2 and A4 by the light of other galaxies,
likely in the lensing cluster. Bottom panel: The
blue-arm spectrum of image B1 of Galaxy B shows a strong
emission line at 4547~\AA, which we interpret as Ly$\alpha$ at $z=2.74$.
Spectra are binned to the spectral resolution.}
\label{fig:shear}
\end{figure}

\section{Gravitational Lens Model}

The geometrical distribution of the lensed images indicates that the projected 
mass distribution
is elliptical in shape and lacks significant substructure.  We have modeled
the mass distribution with a softened power-law ellipsoid 
with surface density $\Sigma \propto
\theta^{-\gamma}$ (Barkana 1998), plus contributions from red cluster members modeled as softened
isothermal spheres and normalized according to their luminosity, 
and additional shear from a sub-group $15''$ to the 
northwest of the BCG, 
modeled as a softened power-law ellipsoid with $\gamma=0.9$.
We find that this simple model, with 15 free parameters
(center of mass coordinates; slope, ellipticity, position angle, core radius, 
normalizations of the cluster, galaxies and subgroup; 
and coordinates of the quasar and of galaxies A and B in the source plane),
and the measured redshifts of galaxies A, B, and the quasar, are
sufficient to fit the two-dimensional positions of 10 lensed 
images (Q1-Q5, A1-A3, B1, B2, i.e., 20 constraints) to a root-mean-square 
accuracy of $\sim0\farcs1$ in the source plane.  

The mass enclosed within a circular aperture including the 
images of galaxy A and corresponding to a radius $\sim 110$~kpc at the cluster redshift,
is $6\times10^{13}~{\rm M}_\odot$. 
We find a mean projected slope of $\gamma
\simeq 0.5$, considerably shallower than isothermal ($\gamma=
1$), but comparable to the projected surface density derived from
CDM-based
 simulations of massive halos (e.g., NFW).
 This result is consistent with an earlier claim of $0.3 <
\gamma < 0.5$ made by Williams \& Saha (2004) using a nonparametric
model based only on the quasar image positions.  
The ellipticity is
relatively large, $e\simeq 0.55$ (suggesting the mass distribution may 
be more complicated than the current model; Edge et al. 2003)
and the mass has a projected position angle of PA$\simeq 332^{\circ}$ (Fig.~1). 
The orientation of the cluster mass distribution is
  approximately aligned with the isophotes of the BCG and the
  position of the sub-group.
In Figure~2, we illustrate how our model
reproduces the morphologies of the lensed images of galaxies A, B, and C.

\begin{deluxetable}{llllll}
\tabletypesize{\scriptsize} 
\tablecaption{Summary of multiply imaged source properties} 
\tablehead{
\colhead{Source} & 
\colhead{No. of}  & 
\colhead{Relative} &
\colhead{predicted $z$} &
\colhead{observed $z$} &
\nl &images & deflection
}

\startdata 
QSO       & 5  & 1.0 & \nodata   & 1.74 \\ 
Galaxy A  & 5  & 1.32 & 3.57   & 3.332\\
Galaxy B  & 2  & 1.255 & 2.65   & 2.74 \\
Galaxy C  & 2  & 1.215 & 2.94   & \nodata \\

\enddata
\end{deluxetable}

\section{Conclusions and Future Work}

We have identified three new multiply imaged galaxies in
ACS images of the cluster lens SDSS J1004+4112, and have obtained redshifts for
two of these systems. The relative deflections of the lensed quasar and
the lensed galaxy images show a clear trend with redshift, increasing 
by $\simeq 30$\% over the range $1.74<z<3.33$. The
model-predicted deflection angles and the measured redshifts are found
to be in good agreement with expectations from a model based on an
elliptical, shallower-than-isothermal, mass distribution and 
the standard cosmological parameters. 

In a future paper (Ofek et al. 2005, in preparation), we will investigate 
what regions of model parameter space can be excluded by the data, in terms of
both the cluster mass distribution and the underlying cosmology, the latter
dictating the ratios of angular diameter distances.
The quality of the strong
lensing constraints available for this cluster should help to break
degeneracies between the parameterization of the cluster mass distribution
and the cosmological world model.  The high redshift of this cluster lens
may also help to reduce uncertainties caused by large scale structure
along the line-of-sight (Dalal et al. 2005).  We will also investigate the
possibility that the mass associated with some of the lensed galaxies
(e.g., the quasar host galaxy) contributes to the lensing signal observed
in more galaxies.  
If such a direct geometric measurement of cosmological parameters 
to a competitive accuracy is feasible, it will apply to a
redshift range intermediate to those based on measurements of 
type-Ia supernovae and the cosmic
microwave background. Furthermore, a lensing-based measurement will 
sample the $\Omega_{m}$ -- $\Omega_{\Lambda}$ parameter plane at an angle
almost parallel to $\Omega_{\Lambda}$ and is therefore complementary
to the two other methods.

Deeper, multi-wavelength imaging of SDSS~J1004+4112 has been
approved, using HST/ACS, HST/NICMOS, and the Spitzer Space Telescope.
These data will likely reveal additional sets of
multiply imaged galaxies, perhaps at even higher redshifts, as well 
as provide reasonably
accurate photometric redshifts, beyond the reach of spectroscopy. Weak
lensing measurements will also be feasible. Together with the
strong lensing information, we will then attempt to tighten considerably the lens
model and to estimate cosmological parameters based on this
uniquely useful lens.

\acknowledgments

We thank Rennan Barkana for his help with the modeling, and Phil Marshall
for his assistance with the Keck observations.
Some of the data presented herein were obtained at the W.M. Keck
 Observatory, which is operated as a scientific partnership among
 Caltech, the University of California and NASA. 
  This work was supported by grant HST-GO-9744 (CSK and EEF)
from the Space Telescope Science Institute
    which is operated by AURA, Inc., under NASA contract NAS5-26555;
by a grant from the German Israeli Foundation for Research and
Development (KS, EOO, and DM);
and by a grant from the Israel Science Foundation (TJB).
GPS acknowledges Caltech Optical Observatories for generously 
supporting his observations of high redshift galaxy
cluster lenses.
%%%%%%%%%%%%%%%%%%%%%%%%%%%%%%%%%%%%%%%%%%%%%%%%%%%%%%%%%%%%%%%%%%%%%%%%%

%%%%%%%%%%%%%%%%%%%%%%%%%%%%%%%%%%%%%%%%%%%%%%%%%%%%%%%%%%%%%%%%%%%%%%

\end{document}